\documentclass[11pt,dvips]{article}
cm \voffset = -2truecm

\usepackage{graphicx}
\usepackage{amssymb}
\usepackage{latexsym}

\begin{document}

\begin{titlepage}

\setcounter{page}{1}
\renewcommand{\thefootnote}{\fnsymbol{footnote}}

\vspace{5mm}
\begin{center}

 {\large \bf Generalized Weyl-Heisenberg algebra, qudit systems and entanglement measure of symmetric states via spin coherent states}

\vspace{1.5cm}

{\bf M. Daoud$^{1,2}${\footnote{email: {\sf m\_daoud@hotmail.com}} and M. R. Kibler$^{2,3,4}${\footnote{email: {\sf m.kibler@ipnl.in2p3.fr} }}}}

\vspace{0.8cm}

{\small $^1$ Department of Physics, Faculty of Sciences Ain Chock, University Hassan II, Casablanca, Morocco \\ 
				 $^2$ CNRS/IN2P3, Institut de Physique Nucl\'eaire, 69622 Villeurbanne, France \\ 
				 $^3$ Facult\'e des Sciences et Technologies, Universit\'e Claude Bernard Lyon 1, 69622 Villeurbanne, France \\ 
         $^4$ Universit\'e de Lyon, 69361 Lyon, France} 

\end{center}

\vspace{1.cm}

\begin{abstract}
A relation is established in the present paper between Dicke states in a $d$-dimensional space and vectors in the representation space of a generalized Weyl-Heisenberg algebra of finite dimension $d$. This provides a natural way to deal with the separable and entangled states of a system of $N = d-1$ symmetric qubit states. Using the decomposition property of Dicke states, it is shown that the separable states coincide with the Perelomov coherent states associated with the generalized Weyl-Heisenberg algebra considered in this paper. In the so-called Majorana scheme, the qudit ($d$-level) states are represented by $N$ points on the Bloch sphere; roughly speaking, it can be said that a qudit (in a $d$-dimensional space) is describable by a $N$-qubit vector (in a $N$-dimensional space). In such a scheme, the permanent of the matrix describing the overlap between the $N$ qubits makes it possible to measure the entanglement between the $N$ qubits forming the qudit. This is confirmed by a Fubini-Study metric analysis. A new parameter, proportional to the permanent and called {\em perma-concurrence}, is introduced for characterizing the entanglement of a symmetric qudit arising from $N$ qubits. For $d=3$ ($\Leftrightarrow N = 2$), this parameter constitutes an alternative to the concurrence for two qubits. Other examples are given for $d=4$ and $5$. A connection between Majorana stars and zeros of a Bargmmann function for qudits closes this article. 
\\
\end{abstract}

\noindent {\em key words: generalized Weyl-Heisenberg algebra; qubit and qudit systems; Dicke states; Perelomov coherent states; entanglement; concurrence; perma-concurrence; Fubini-Study metric; Majorana stars; Bargmann function} 
\\ 

\end{titlepage}

\section{Introduction}

Geometrical representations are of particular interest in various problems of quantum mechanics. For instance, the Bloch representation is widely used in the context
of characterizing quantum correlations in multiqubit systems \cite{N-Bloch,N-Bloch2,qudit-Bloch}. This representation is based on the idea of Majorana to visualize a $j$-spin as a set of $2j$ points in a sphere \cite{Majorana}. The Bloch sphere was used in the study of entanglement quantification and classification in multiqubit systems \cite{Dur.Vid.Cir:00,Class3,Class}. The investigation and the understanding of quantum correlations in multipartite quantum systems are essential in several branches of quantum information such as quantum cryptography \cite{Crypto}, quantum teleportation \cite{Teleportation} and quantum 
communication \cite{Communication,Communication2}. 
\\

The separability for two-qubit states can be addressed with the concept of the Wootters concurrence \cite{Concurrence1,Concurrence}. However, for multiqubit quantum systems, the measure of quantum correlations is very challenging. Several ways to understand the main features of entangled multiqubit states were employed in the literature \cite{Tangle,Barycenter}. 
\\

Algebraic and geometrical methods were intensively used in quantum mechanics \cite{Ana:91,Sch:96,Ash.Sch:99,Bro.Hug:00} and continue nowadays to contribute to our understanding of entanglement properties in multipartite quantum systems (for instance, see \cite{Kus.Zyc:01,Ben.Bra.Zyc:02,Bro.Gus.Hug:07,Licata}). In this spirit, several works were devoted to geometrical analysis of entangled multipartite states to find the best measure to quantify the amount of entanglement in a multiqubit system  \cite{Dur.Vid.Cir:00,Kus.Zyc:01,Mos.Dan:01,Wei.Gol:03,Hub.Kle.Wei.Gon.Guh:09,Aulbach2,Mar.Gir.Bra.Bra.Bas:10,Che.Aul.Haj:14,Bag.Bas.Mar:14}. The classification of multipartite entangled states was investigated from several perspectives using different geometrical tools \cite{Miy:03,Hey:08,Hol.Luq.Thi:12} to provide the appropriate way  to approach the quantum correlations in multiqubit states. Among these quantum states, $j$-spin coherent states are of special interest \cite{Rad:71}. Indeed, they are the most classical (in contrast to quantum) states and can be viewed as $2j$-qubit states which are completely separable. In this sense, spin coherent states can be used to characterize the entanglement in totally symmetric multiqubit systems \cite{Man.Cou.Kel.Mil:14}. 
\\

The multipartite quantum states, invariant under permutation symmetry, have attracted a considerable attention during the last decade. This is essentially motivated by their occurrence in the context of multipartite entanglement \cite{Class,Aulbach2,Stockton1,Mathonet,Markham,Stockton3,Aulbach1,Stockton2} and quantum tomography \cite{Toth1,Toth2,Toth3}. In fact, the dimension $2^N$ of the Hilbert space for an ensemble of $N$ qubits system reduces to $N+1$ when the whole system possesses the exchange symmetry. The appropriate representations to deal with the totally symmetric states are the Dicke basis \cite{Dicke,Toth2007,Bergmann2013} and Majorana 
representation \cite{Majorana}. 
\\

In the present paper, we consider a realization of the generalized Weyl-Heisenberg algebra, introduced in \cite{Dao-Kib2006,Dao-Kib2010,Dao-Kib2011}, by means of an ensemble of two-qubit operators. We investigate the correspondence between the vectors of the representation space of the Weyl-Heisenberg algebra and the Dicke states. Using the decomposition properties of Dicke states, we show that the separable states are necessarily the Perelomov coherent states associated with the generalized Weyl-Heisenberg algebra. The coherent states are written as tensor products of single qubit coherent states. We also discuss the separability in terms of the permanent of the matrix of the overlap between spin coherent states. 
\\

\section{Qubits and generalized Weyl-Heisenberg algebra}

\subsection{Bosonic and fermionic algebras}

The study of bosonic and fermionic many particle states is simplified  by considering the algebraic structures of the corresponding raising and lowering operators. On the one hand, for bosons the creation operators $b_i^+$ and the annihilation operators $b_i^-$ satisfy the commutation relations
\begin{eqnarray}\label{CR}
 [ b_i^- , b_j^+ ] = \delta_{ij} \mathbb{I}, \quad [  b_i^- , b_j^- ] = [  b_i^+  , b_j^+ ] = 0,
\end{eqnarray}
where $\mathbb{I}$ stands for the identity operator. On the other hand, fermions are specified by the following anti-commutation relations
\begin{eqnarray}\label{ACR}
\{  f_i^- , f_j^+  \} = \delta_{ij} \mathbb{I}, \quad \{  f_i^+ , f_j^+  \} = \{  f_i^- , f_j^-  \} = 0
\end{eqnarray}
of the creation operators $f_i^+$ and the annihilation operators $f_i^-$. The properties of Fock states follow from the commutation and anti-commutation relations which impose only one particle in each state for fermions (in a two-dimensional space) and an arbitrary number of particles for bosons (in an infinite-dimensional space). Following Wu and Lidar \cite{Wu-Lidar}, there is a crucial difference between fermions and qubits (two level systems). In fact, a qubit  is a vector in a 
two-dimensional Hilbert space as for fermions and the Hilbert space of a multiqubit system has a tensor product structure like for bosons. In this respect, the raising and lowering operators commutation rules for qubits are neither specified by relations of bosonic type (\ref{CR}) nor of fermionic type (\ref{ACR}).

\subsection{Qubit algebra}

The algebraic structure relations for qubits are different from those defining Fermi and Bose operators. Indeed, denoting by $\vert 0 \rangle$ and $\vert 1 \rangle$ the  states of a two-level system (qubit), the lowering ($q^-$), raising ($q^+$), and number ($K$) operators defined by
		\begin{eqnarray}
q^- = |0 \rangle \langle 1|           &\Rightarrow& q^- |1 \rangle = |0 \rangle,  \quad q^- |0 \rangle = 0 
		\label{1def qplus et qmoins} \\
q^+ = |1 \rangle \langle 0|           &\Rightarrow& q^+ |0 \rangle = |1 \rangle,  \quad q^+ |1 \rangle = 0 
		\label{2def qplus et qmoins} \\
K   = \vert 1 \rangle \langle 1 \vert &\Rightarrow& K |1 \rangle = |1 \rangle,  \quad K |0 \rangle = 0 
		\label{3def qplus et qmoins}
		\end{eqnarray}
satisfy the relations
\begin{eqnarray}\label{kappamoinsun}
(q^-)^{\dagger} = q^+, \quad K^{\dagger} = K, \quad [q^- , q^+] = \mathbb{I} - 2 K, \quad [K , q^+] = + q^+, \quad [K , q^-] = - q^-
\end{eqnarray}
(we use $A^{\dagger}$ to denote the adjoint of $A$). Furthermore, the creation and the annihilation
operators satisfy the nilpotency conditions
\begin{eqnarray}
(q^+)^2 = (q^-)^2 = 0,
\label{q-nilp}
\end{eqnarray}
as in the case of fermions. 
\\ 

We note that the commutation relations in (\ref{kappamoinsun}) coincide with those defining the algebra introduced in \cite{andrzej} to provide an alternative algebraic description of qubits instead of the parafermionic formulation considered in \cite{Wu-Lidar}. In addition, the generalized oscillator algebra 
${\cal A}_{\kappa}$ introduced in \cite{Dao-Kib2010} as a particular case of the generalized Weyl-Heisenberg algebra of bosonic type \cite{Dao-Kib2006} provides an alternative description of qubits 
(in \cite{Dao-Kib2011}, the algebra ${\cal A}_{\kappa}$ is also denoted as ${\cal A}_{\kappa}(1)$ in view of its extension to ${\cal A}_{\kappa}(2)$). In fact, Eqs.~(\ref{kappamoinsun}) correspond to $\kappa = -1$. 

\subsection{Qudit algebra}

To give an algebraic description  of  $d$-dimensional quantum systems $(d \geq 2)$, we consider a set of $N = d-1$ qubits. We denote as $q^+_i$, $q^-_i$, and 
$K_i$  the raising, lowering, and number operators associated with the $i$-th qubit. They satisfy relations similar to (\ref{kappamoinsun}), namely, 
\begin{eqnarray}\label{RC-qubits}
(q^-_i)^{\dagger} = q^+_i, \quad (K_i)^{\dagger} = K_i, \quad[q^-_i , q^+_j] = (\mathbb{I} - 2 K_i)\delta_{ij}, \quad 
[K_i , q^+_j] = + \delta_{ij} q^+_i, \quad [K_i , q^-_j] = - \delta_{ij} q^-_i 
\end{eqnarray}
and 
\begin{eqnarray}
[q^-_i , q^-_j] = [q^+_i , q^+_j] = 0
\label{RC-qubits-bis}
\end{eqnarray}
for $i, j = 1, 2, \cdots, N$.
\\

Let us denote as ${\cal H}_2$ the two-dimensional Hilbert space for a single qubit. An orthonormal basis of ${\cal H}_2$ is given by the set 
$$
\{  \vert n \rangle : n = 0, 1  \}.
$$ 
The multiqubit $2^N$-dimensional Hilbert space ${\cal H}_{2^N}$ for the $N$ qubits has the following  tensor product structure
$$ 
{\cal H}_{2^N} = {\cal H}_2 \otimes {\cal H}_2 \otimes \cdots \otimes {\cal H}_2
$$
(with $N \geq 1$ factors), like for bosons. In other words, the set 
			$$
			\{ | n_1 n_2 \cdots n_N \rangle : n_i = 0,1 \ (i= 1, 2, \cdots, N) \}, 
			$$
where 
$$
| n_1 n_2 \cdots n_N \rangle = | n_1 \rangle \otimes | n_2 \rangle \otimes \cdots \otimes| n_N \rangle,
$$ 
constitutes an orthonormal basis of ${\cal H}_{2^N}$. The Dicke states shall be defined in Section \ref{Dicke states} as linear combinations of the states $| n_1 n_2 \cdots n_N \rangle$. 
\\

We define the collective lowering, raising and number operators in the Hilbert space  ${\cal H}_{2^N}$ as follows
\begin{eqnarray}
q^- = \sum_{i=1}^{N} q_i^-,  \quad q^+ = \sum_{i=1}^{N} q_i^+, \quad K = \sum_{i=1}^{N} K_i
\label{green}
\end{eqnarray}
in terms of the annihilation, creation, and number operators $q_i^- $, $q_i^+$, and $K_i$. In Eq.~(\ref{green}), $q_i^{\pm}$ should be understood as the operator $\mathbb{I} \otimes \cdots \otimes \mathbb{I} \otimes q_i^{\pm} \otimes \mathbb{I} \otimes \cdots \otimes \mathbb{I}$, where $q_i^{\pm}$ stands, among the $N$ operators, at the $i$-th position from the left. It is trivial to check that 
$$
q^- | 0 0 \cdots 0 \rangle = 0, \quad q^+ | 1 1 \cdots 1 \rangle = 0. 
$$
The action of $q^-$ and $q^+$ on vectors $| n_1 n_2 \cdots n_N \rangle$ involving qubits $| 0 \rangle$ and $| 1 \rangle$, as for Dicke states, shall be considered in Section \ref{Dicke states}.  
\\

By using Eqs.~(\ref{q-nilp}), (\ref{RC-qubits-bis}), and (\ref{green}), we obtain
\begin{eqnarray}\label{power-q}
(q^-)^{k} =  k! \sum_{i_1< i_2< \cdots<i_k} q_{i_1}^- q_{i_2}^- \cdots q_{i_k}^-,  \quad 
(q^+)^{k} =  k! \sum_{i_1< i_2< \cdots<i_k} q_{i_1}^+ q_{i_2}^+ \cdots q_{i_k}^+
\end{eqnarray}
for $k = 1, 2, \cdots, N$. In particular, for $k = N$, the relations (\ref{power-q}) give
$$
(q^-)^{N} =  N! q_{1}^- q_{2}^- \cdots q_{N}^-,  \quad 
(q^+)^{N} =  N! q_{1}^+ q_{2}^+ \cdots q_{N}^+,  
$$
which lead to the  nilpotency relations
\begin{eqnarray}\label{nilpo1}
(q^-)^{N+1} = (q^+)^{N+1}  = 0.
\end{eqnarray}
Equation (\ref{nilpo1}) for $N = 1$ ($\Leftrightarrow d = 2$) gives back Eq.~(\ref{q-nilp}) which is reminiscent of the Pauli exclusion principle for fermions. 
\\

In view of (\ref{RC-qubits}) and (\ref{green}), the qudit operators $q^+$, $q^-$, and $K$ satisfy the commutation rules
\begin{eqnarray}\label{qudit}
[ q^- , q^+ ] = N \mathbb{I} - 2K, \quad   [K , q^+] = + q^+, \quad   [K , q^-] =  - q^-,
\end{eqnarray}
which are similar to the relations defining the generalized Weyl-Heisenberg algebra ${\cal A}_{\kappa}$ introduced in \cite{Dao-Kib2010}. More precisely, let us put 
	\begin{eqnarray}
a^{\pm} = \frac{1}{\sqrt{N}} q^{\pm}.
 	\label{10bis}
	\end{eqnarray}
Then, we have the relations
	\begin{eqnarray}
[a^- , a^+] = \mathbb{I} + 2 \kappa K, \quad 
 [K , a^{\pm}] = \pm a^{\pm}, \quad 
         (a^-)^\dagger = a^+, \quad K^\dagger = K,
 	\label{10ter}
	\end{eqnarray}
where the parameter $\kappa$ is 
	\begin{eqnarray}
\kappa = - \frac{1}{N}.
	\label{10quart}
 	\end{eqnarray}
Therefore, the operators $a^-$, $a^+$, and $K$ generate the algebra ${\cal A}_{\kappa}$ with 
$\kappa = - \frac{1}{N}$. This shows that the algebra ${\cal A}_{\kappa}$ can be described by a set of $N$ qubits. According to the analysis in \cite{Dao-Kib2010}, since $- \frac{1}{N} < 0$, the algebra 
${\cal A}_{\kappa}$ admits finite-dimensional representations. Indeed, we shall show that the representation constructed on the basis $\{ | N ; k \rangle : k = 0, 1, \cdots, N \}$ of the Dicke states 
(see Section \ref{Dicke states}) is of dimension $d = N+1$. 
\\

Note that the lowering and raising operators $q^+$ and $q^-$ close the following trilinear commutation relations
$$
[q^-, [ q^+ , q^- ]] = + 2 q^-, \quad  
[q^+, [ q^+ , q^- ]] = - 2 q^+ 
$$
like in a para-fermionic algebra \cite{Palev1}. Note also that the definition (\ref{green}) is identical to the decomposition used by Green for defining para-fermions from ordinary fermions \cite{refgreen}. 

\section{ Dicke states} \label{Dicke states}

\subsection{Definitions}

The Hilbert space ${\cal H}_{2^N}$ can be partitioned as 
			\begin{eqnarray}
			{\cal H}_{2^N} = \bigoplus_{k = 0}^N {\cal F}_{N,k}, 
			\label{decompositionH2N}
			\end{eqnarray}
where the sub-space ${\cal F}_{N,k}$ is spanned by the orthonormal set 
			\begin{eqnarray*}
			\{ | n_1 n_2 \cdots n_N \rangle : n_1 + n_2 + \cdots + n_N = k \}. 
			\end{eqnarray*}
Each vector $| n_1 n_2 \cdots n_N \rangle$ of ${\cal F}_{N,k}$ contains $N-k$ qubits $| 0 \rangle$ and $k$ qubits $| 1 \rangle$. The dimension of the space 
${\cal F}_{N,k}$ is given by 
      \begin{eqnarray*}
			\dim {\cal F}_{N,k} = C_N^k = \frac{N!}{k! (N-k)!}, 
			\end{eqnarray*}	
in terms of the binomial coefficient $C_N^k$, and satisfies 
			\begin{eqnarray*}
			\dim {\cal H}_{2^N} = \sum_{k=0}^N \dim {\cal F}_{N,k} = 2^N. 
			\end{eqnarray*}	
Clearly, ${\cal F}_{N,k}$ is invariant under any of the $N!$ permutations of the $N$ qubits. The orthogonal decomposition (\ref{decompositionH2N}) of 
${\cal H}_{2^N}$ turns out to be useful in the definition of Dicke states.  
\\		

To each ${\cal F}_{N,k}$ it is possible to associate a Dicke state $| N ; k \rangle$ which is the sum (up to a normalization factor) of the various states of 
${\cal F}_{N,k}$. To be more precise, let us define the Dicke state $| N ; k \rangle$ as follows \cite{Toth2007,Bergmann2013}
 			\begin{eqnarray}
|N ; k\rangle = \sqrt{\frac{k! (N-k)!}{N!}} \sum_{ \{ \sigma \} } 
\sigma | \underbrace{0 0 \cdots 0}_{N-k} \underbrace{11 \cdots 1}_{k} \rangle, \quad 0 \leq k \leq N, 
			\label{Dickestate}
			\end{eqnarray}		
where the number of $0$ and $1$ in the vector $|0 0 \cdots 011 \cdots 1\rangle$ are $N-k$ and $k$, respectively. Furthermore, the summation over $\{ \sigma \}$ runs on the permutations $\sigma$ of the symmetric group $S_N$ restricted to the identity permutation and the permutations between the $0$'s 
and $1$'s (the permutations between the various $0$'s as well as those between the various $1$'s are excluded, only the permutations between the $0$'s and $1$'s leading to distinct vectors are permitted). Each vector in (\ref{Dickestate}) involves $(N-k)+k = N$ qubits. A Dicke state $|N ; k\rangle$ is thus a normalized symmetrical superposition of the states of ${\cal F}_{N,k}$. More precisely, 
Eq.~(\ref{Dickestate}) means 
 			\begin{eqnarray*}
|N ; k\rangle = \sqrt{\frac{k! (N-k)!}{N!}} \sum_{ | x \rangle \in {\cal F}_{N,k} } | x \rangle.			
			\end{eqnarray*}		
Indeed, each Dicke state $| N ; k \rangle$ and, more  generally, any linear combination of the $N+1$ Dicke states $| N ; k \rangle$ (with $k = 0, 1, \cdots, N$) transform as the totally symmetric irreducible representation $[N]$ of the group $S_N$ of the permutations of the $N = d-1$ qubits. 
\\

As a trivial example, for $N=1$ the Dicke states $| 1 ; 0 \rangle$ and $| 1 ; 1 \rangle$ are nothing but the one-qubit states $| 0 \rangle$ and $| 1 \rangle$, respectively (these qubit states are generally associated with the angular momentum states $|\frac{1}{2} , \frac{1}{2})$ and $|\frac{1}{2} , -\frac{1}{2})$, respectively). As a more instructive example, for $N = 4$ we have the $d=N+1=5$ Dicke states 
	\begin{eqnarray}
|4 ; 0\rangle &=& |0000\rangle, \nonumber \\
|4 ; 1\rangle &=& \frac{1}{2} (|0001\rangle + |0010\rangle + |0100\rangle + |1000 \rangle), \nonumber                                          \\
|4 ; 2\rangle &=& \frac{1}{\sqrt{6}} (|0011\rangle + |0101\rangle + |0110\rangle + |1001\rangle + |1010\rangle + |1100\rangle), \nonumber        \\  
|4 ; 3\rangle &=& \frac{1}{2} (|0111\rangle + |1011\rangle + |1101\rangle + |1110 \rangle), \nonumber                                          \\
|4 ; 4\rangle &=& |1111\rangle. \nonumber
 	\end{eqnarray}
Each vector $|4 ; k\rangle$ is a symmetric (with respect to $S_4$) linear combination of the vectors of ${\cal F}_{4,k}$. 
\\

For fixed $N$, we have
	\begin{eqnarray*}
\langle N;k | N;\ell \rangle = \delta_{k,\ell}, \quad k,\ell = 0, 1, \cdots, N, 
 	\end{eqnarray*}
so that the set $\{ |N ; k\rangle : k = 0, 1, \cdots, N \}$ constitutes an orthonormal system in the space ${\cal H}_{2^N}$. Let us denote as ${\cal G}_{d}$ the space of dimension $d = N+1$ spanned by the $N+1$ symmetric vectors $|N ; k\rangle$ with $k = 0, 1, \cdots, N$. Then, the set 
$\{ |N ; k\rangle : k = 0, 1, \cdots, N \}$ is an orthonormal basis of ${\cal G}_{d}$. 

\subsection{Dicke states and representations of ${\cal A}_{\kappa}$}

The nilpotency relations (\ref{nilpo1}) imply that  the representation space of the generalized Weyl-Heisenberg algebra ${\cal A}_{\kappa}$ with 
$\kappa = - \frac{1}{N}$, see Eqs.~(\ref{qudit})-(\ref{10quart}), is of dimension  $d = N+1$. The representation vectors can be determined using repeated actions of the raising operator $q^+$ combined with the actions of the operators $q_i^-$ and $q_i^+$ defined by relations similar to 
(\ref{1def qplus et qmoins})-(\ref{3def qplus et qmoins}). It can be shown that these representation vectors are Dicke states. The proof is as follows.  
\\

First, the action of the operator $q^+$ on the ground state $\vert 0 0 \cdots 0 \rangle$ of ${\cal G}_{d}$ yields 
\begin{eqnarray*}
 q^+ \vert 0 0 \cdots 0 \rangle = \vert 1 0 \cdots 0 \rangle + \vert 0 1 \cdots 0 \rangle + \cdots + \vert 0 0 \cdots 1 \rangle
\end{eqnarray*}
or equivalently 
	\begin{eqnarray}
	q^+ \vert N ; 0 \rangle = \sqrt{N} \vert N ; 1 \rangle. 
	\label{q11a}
	\end{eqnarray}
Second, the action of $(q^+)^2$ on $\vert 0 0 \cdots 0 \rangle$ gives
\begin{eqnarray*}
(q^+)^2 \vert 0 0 \cdots 0 \rangle = 2 \left( \vert 1 1 0 \cdots 0 \rangle + \vert 1 0 1 \cdots 0 \rangle + \cdots + \vert 0 0 \cdots 0 1 1 \rangle \right)
\end{eqnarray*}
or 
\begin{eqnarray*}
(q^+)^2 \vert N ; 0 \rangle = \sqrt{2N(N-1)} \vert N ; 2 \rangle. 
\end{eqnarray*}
From repeated application of the raising operator $q^+$ on the state $\vert 0 0 \cdots 0 \rangle$ of 
${\cal G}_{d}$, we obtain 
\begin{eqnarray}
(q^+)^k \vert N ; 0 \rangle = \sqrt{\frac{k! N!}{(N-k)!}} \vert N ; k \rangle.
\label{equ3}
\end{eqnarray}
By using Eq.~(\ref{equ3}), we finally get the ladder relation 
\begin{eqnarray}
q^+ \vert N ; k \rangle = \sqrt{(k+1)(N - k)} \vert N ; k+1 \rangle.
\label{BB}
\end{eqnarray}
Similarly, for the lowering operator $q^-$, we have  
\begin{eqnarray}
q^- \vert N ; k \rangle = \sqrt{k(N-k + 1)} \vert N ; k-1 \rangle. 
\label{CC}
\end{eqnarray}
Equations (\ref{BB}) and (\ref{CC}) can be rewritten as 
	\begin{eqnarray}
q^+ | N ; k \rangle &=& \sqrt{F(N , k + s + \frac{1}{2})} | N ; k+1 \rangle, \label{BBCC1} \\ 
q^- | N ; k \rangle &=& \sqrt{F(N , k + s - \frac{1}{2})} | N ; k-1 \rangle, \label{BBCC2}
	\end{eqnarray}
where $s = \frac{1}{2}$ and 
	\begin{eqnarray*}
F(N , \ell ) = \ell (N - \ell + 1), \quad 0 \leq \ell \leq N + 1. 
 	\end{eqnarray*}
Note that 	
\begin{eqnarray*} 
q^+ | N ; N \rangle = q^- | N ; 0 \rangle = 0
\end{eqnarray*}
gives the action of the raising and lowering qubit operators on the extremal Dicke states $\vert N ; N \rangle$ and $\vert N ; 0 \rangle$ of ${\cal G}_{d}$. 
\\

The Dicke states are eigenstates of the operator $K$ defined in (\ref{green}), see also (\ref{3def qplus et qmoins}) in the case $N=1$. Indeed, we have  
	\begin{eqnarray}
K | N ; k \rangle = k | N ; k \rangle, \quad k = 0, 1, \cdots, N,
 	\label{DD}
	\end{eqnarray}
in agreement with the fact that $K$ is a number operator: it counts the number of qubits of type $| 1 \rangle$ in the Dicke state $| N ; k \rangle$. From 
Eqs.~(\ref{BBCC1}), (\ref{BBCC2}), and (\ref{DD}), we recover the commutation relations (\ref{qudit}). Therefore, the generalized Weyl-Heisenberg algebra 
${\cal A}_{\kappa}$, with $\kappa = - \frac{1}{N}$, generated by the operators $\frac{1}{\sqrt{N}}q^+$, 
$\frac{1}{\sqrt{N}}q^-$, $K$, and $\mathbb{I}$ provides an algebraic  description of a qudit ($d$-level) system viewed as a collection of $N = d-1$ qubits. As a matter of fact, the vectors of the representation space ${\cal G}_{d}$ of  the algebra ${\cal A}_{\kappa}$ are the Dicke states $| N ; k \rangle$ which are symmetric superpositions of states of a multiqubit system.

\subsection{ Decomposition of Dicke states}

Let us consider again the action of $q^+$ on the ground state $\vert 0 0 \cdots 0 \rangle$ involving $N$ qubits $| 0 \rangle$. We have seen that 
	\begin{eqnarray}
	q^+ \vert 0 0 \cdots 0 \rangle = q^+ \vert N ; 0 \rangle = \sqrt{N} \vert N ; 1 \rangle, 
	\label{form1}
	\end{eqnarray}
see Eq.~(\ref{q11a}). On another side, we have 
$$ 
q^+ \vert 0 0 \cdots 0 \rangle = (q_1^+ + q_2^+ + \cdots + q_{N}^+) \vert 0 0 \cdots 0 \rangle.
$$
This gives
$$ 
q^+ \vert 0 0 \cdots 0 \rangle = \vert 1 0 \cdots 0 \rangle +  
                                 \vert 0 1 \cdots 0 \rangle + \cdots + 
																 \vert 0 0 \cdots 1 \rangle,  
$$
which can be rewritten as 
$$ 
q^+ \vert 0 0 \cdots 0\rangle = \left( 
\vert 1 0 \cdots 0 \rangle + 
\vert 0 1 \cdots 0 \rangle + \cdots + 
\vert 0 0 \cdots 1 \rangle \right) \otimes \vert 0\rangle + 
\vert 0 0 \cdots 0 \rangle         \otimes \vert 1\rangle, 
$$
where the states $| \times \times \cdots \times \rangle$ on the right-hand side member contains $N-1$ qubits. Thus, we get
	\begin{eqnarray}
	q^+ \vert 0 0 \cdots 0 \rangle = \sqrt{N-1} \vert N-1 ; 1 \rangle \otimes \vert 0 \rangle
																					 +  \vert N-1 ; 0 \rangle \otimes \vert 1 \rangle. 
	\label{form2}																			
	\end{eqnarray}
A comparison of (\ref{form1}) and (\ref{form2}) yields 
	\begin{eqnarray}
	\sqrt{N} \vert N ; 1 \rangle = \sqrt{N-1} \vert N-1 ; 1 \rangle \otimes \vert 0 \rangle
																					+ \vert N-1 ; 0 \rangle \otimes \vert 1 \rangle.
	\label{decomp(N,1)}																			
	\end{eqnarray}
By applying the creation operator $q^+$ on both sides of (\ref{decomp(N,1)}) and by using (\ref{BB}), we obtain
\begin{eqnarray*}
\sqrt{2N(N-1)} \vert N ; 2 \rangle = \sqrt{2(N-1)(N-2)} \vert N-1 ; 2 \rangle \otimes \vert 0 \rangle 
                                   + 2 \sqrt{N-1}      \vert N-1 ; 1 \rangle \otimes \vert 1 \rangle. 
\end{eqnarray*}
Repeating this process $k$ times, we end up with 
\begin{eqnarray}
\sqrt{\frac{N!}{k!(N-k)!}} \vert N ; k \rangle &=& \sqrt{\frac{(N-1)!}{k!(N-k-1)!}}   \vert N-1 ; k   \rangle \otimes \vert 0 \rangle  \nonumber \\
                                               &+& \sqrt{\frac{(N-1)!}{(k-1)!(N-k)!}} \vert N-1 ; k-1 \rangle \otimes \vert 1 \rangle. \label{decompo-sep}
\end{eqnarray}
Equation (\ref{decompo-sep}) can be simplified to give 
	\begin{eqnarray}
| N ; k \rangle = \sqrt{\frac{N-k}{N}} | N - 1 ; k   \rangle \otimes | 0 \rangle + 
                  \sqrt{\frac{k  }{N}} | N - 1 ; k-1 \rangle \otimes | 1 \rangle,          \label{somme} 
 	\end{eqnarray}
where $0 \leq k \leq N$. 
\\

For $k \not= 0$ and $k \not= N$, there are two terms in the decomposition of $| N ; k \rangle$: one is a tensor product involving the qubit $| 0 \rangle$ and the other a tensor product involving the qubit $| 1 \rangle$. The decomposition (\ref{somme}) of the Dicke states $| N ; k \rangle$ is trivial in the cases $k = 0$ and $k = N$. For $k = 0$ and $k = N$, the significance of (\ref{somme}) is clear. These two particular cases correspond to a factorization of the Dicke state $| N ; k \rangle$ for $N$ qubits into the tensor product of a Dicke state for $N - 1$ qubits with a state for one qubit. 

\subsection{Dicke states and angular momentum states}

To close this section, a link between Dicke states and angular momentum states is in order. The Lie 
algebra su(2) of the group SU(2) can be realized by means of the angular momentum operators $J_+$, $J_-$, and $J_z$. The irreducible representation $(j)$ of su(2) can be constructed from the set $\{ |j , m) : m = -j, -j+1, \cdots, j\}$ of angular momentum states. We know that 
	\begin{eqnarray}
	J_+ | j , m ) &=& \sqrt{(j - m) (j + m + 1)} | j , m+1 ),  \label{CO1} \\
	J_- | j , m ) &=& \sqrt{(j + m) (j - m + 1)} | j , m-1 ),  \label{CO2} \\
	J_z | j , m ) &=& m | j , m ),  \label{CO3} 
	\end{eqnarray}
according to the Condon and Shortley phase convention \cite{CondonOdaba}. Let us put 
	\begin{eqnarray*}
	k = j - m, \quad N - k = j + m \quad \Leftrightarrow \quad j = \frac{N}{2}, \quad m = -k + \frac{N}{2}.
	\end{eqnarray*}
Therefore, the state $| j , m )$ can be denoted as $| N ; k \rangle$ since, for fixed $j$, then $N$ is fixed and $-j \leq m \leq j$ implies $0 \leq k \leq N$. Consequently, Eqs.~(\ref{CO1})-(\ref{CO3}) can be rewritten as 
	\begin{eqnarray*}
	J_+ | N ; k \rangle &=& \sqrt{k (N - k + 1)} | N ; k-1 \rangle,   \\
	J_- | N ; k \rangle &=& \sqrt{(N-k) (k + 1)} | N ; k+1 \rangle,   \\
	J_z | N ; k \rangle &=& \left( \frac{N}{2} - k \right) | N ; k \rangle,  
	\end{eqnarray*}
to be compared with Eqs.~(\ref{BB}), (\ref{CC}), and (\ref{DD}). This leads to the identification
	\begin{eqnarray*}
	J_+ = q^-, \quad 
	J_- = q^+, \quad
	J_z = \frac{N}{2} I - K
	\end{eqnarray*}
which establishes a link between the Weyl-Heisenberg algebra $A_{\kappa}$ with $\kappa = -\frac{1}{N}$ and the Lie algebra su(2). The rewriting of the Dicke state 
$| N ; k \rangle$, see Eq.~(\ref{Dickestate}), in terms of the variables $j$ and $m$ yields
			\begin{eqnarray*}
&& |j , m ) = \sqrt{\frac{(j - m)! (j + m)!}{(2j)!}} \\ 
&& \times \sum_{ \{ \sigma \} } \sigma \left(
\underbrace{|\frac{1}{2} , \frac{1}{2}) \otimes |\frac{1}{2} , \frac{1}{2}) \otimes \cdots \otimes |\frac{1}{2} , \frac{1}{2})}_{j+m} \otimes 
\underbrace{|\frac{1}{2} ,-\frac{1}{2}) \otimes |\frac{1}{2} ,-\frac{1}{2}) \otimes \cdots \otimes |\frac{1}{2} ,-\frac{1}{2})}_{j-m} \right), 
			\end{eqnarray*}		
where the summation over $\{ \sigma \}$ runs on the permutations $\sigma$ of the symmetric group $S_{2j}$ restricted to the identity permutation and the permutations between the states 
$|\frac{1}{2} ,  \frac{1}{2})$ and $|\frac{1}{2} , -\frac{1}{2})$ exclusively (only the permutations leading to distinct vectors are permitted).

\section{ Separable qudit states}

\subsection{Factorization of a qudit}
 
In this section, we start from a qudit ($d$-level state) and study on which condition such a state is separable in the direct product of $d-1$ qubit states.
\\

The most general state in the space ${\cal G}_{d}$ can be considered as a qudit $| \psi_d \rangle$ constituted from $N = d-1$ qubits. In other words, in terms of Dicke states we have 
	\begin{eqnarray}
| \psi_d \rangle = \sum_{k=0}^N c_k | N ; k \rangle, \quad N = d - 1, \quad c_k \in \mathbb{C}. 
	\label{psiN} 
 	\end{eqnarray}
We may ask the question: on which condition the vector $| \psi_d \rangle$ can be factorized as 
	\begin{eqnarray*}
| \psi_d \rangle = | \phi_{d-1} \rangle \otimes | \varphi_1 \rangle
 	\end{eqnarray*}
involving a state $| \phi_{d-1} \rangle$ for $N-1$ qubits and a state $| \varphi_1 \rangle$ for one qubit?
\\

The use of Eq.~(\ref{somme}) yields 
	\begin{eqnarray*}
| \psi_d \rangle = \sum_{k=0}^N c_k 
\left[ \sqrt{\frac{N-k}{N}} | N - 1 ; k   \rangle \otimes | 0 \rangle + 
                  \sqrt{\frac{k  }{N}} | N - 1 ; k-1 \rangle \otimes | 1 \rangle \right], 
 	\end{eqnarray*}
which can be rewritten as 
	\begin{eqnarray*}
| \psi_d \rangle = | u \rangle \otimes | 0 \rangle + 
                   | v \rangle \otimes | 1 \rangle, 
 	\end{eqnarray*}
where 
	\begin{eqnarray}
| u \rangle = \sum_{k=0}^{N-1} c_k \sqrt{\frac{N-k}{N}} | N - 1 ; k   \rangle, \quad      
| v \rangle = \sum_{k=1}^N     c_k \sqrt{\frac{k  }{N}} | N - 1 ; k-1 \rangle. 
\label{uN-1}
 	\end{eqnarray}
Clearly, the state $| \psi_d \rangle$ is separable if there exists $z$ in $\mathbb{C}$ such that 
	\begin{eqnarray}
| v \rangle = z | u \rangle.     
\label{proportion:v-u}
 	\end{eqnarray}
Then 
	\begin{eqnarray}
| \psi_d \rangle  = | u \rangle \otimes (| 0 \rangle + z | 1 \rangle),   
\label{psiNproduitde2}
 	\end{eqnarray}
where
	\begin{eqnarray*}
| u \rangle \equiv | \phi_{d-1} \rangle, \quad | 0 \rangle + z | 1 \rangle \equiv | \varphi_1 \rangle.
 	\end{eqnarray*}
It is easy to show that Eq.~(\ref{proportion:v-u}) implies
	\begin{eqnarray*}
  \sum_{k=0}^{N-1} c_{k+1} \sqrt{k+1} | N - 1 ; k \rangle = 
z \sum_{k=0}^{N-1} c_k     \sqrt{N-k} | N - 1 ; k \rangle. 
	\end{eqnarray*}
Consequently, we get the recurrence relation 
	\begin{eqnarray*}
z c_k \sqrt{N-k} = c_{k+1} \sqrt{k+1}
	\end{eqnarray*}
that admits the solution 
	\begin{eqnarray*}
c_k = c_0 z^k \sqrt{C_N^k}, 
	\end{eqnarray*}
where the coefficient $c_0$ can be calculated from the normalization condition 
$\langle \psi_d | \psi_d \rangle = 1$. This leads to
	\begin{eqnarray}
c_k = \frac{z^k}{(1 + {\bar z} z)^{\frac{N}{2}}} \sqrt{\frac{N!}{k! (N-k)!}}, \quad k = 0, 1, \cdots, N    
\label{ck}
	\end{eqnarray}
up to a phase factor. Thus, the introduction of (\ref{ck}) into (\ref{psiN}) leads to the separable state 
	\begin{eqnarray}
| \psi_d \rangle = \frac{1}{(1 + {\bar z} z)^{\frac{N}{2}}} \sum_{k=0}^N z^k \sqrt{\frac{N!}{k! (N-k)!}} | N ; k \rangle.      
\label{psiNseparable}
	\end{eqnarray} 
In order to identify the various factors occurring in the decomposition of the separable state (\ref{psiNseparable}), as a tensor product, we note that the use of (\ref{ck}) in (\ref{uN-1}) gives 
	\begin{eqnarray*}
| u \rangle = \frac{1}{\sqrt{1 + {\bar z} z}} | \psi_{d-1} \rangle.
	\end{eqnarray*}
Hence, Eq.~(\ref{psiNproduitde2}) takes the form 
	\begin{eqnarray}
| \psi_d \rangle = | \psi_{d-1} \rangle \otimes | z \rangle,      
\label{psiNpsiN-1}
	\end{eqnarray}
where 
	\begin{eqnarray}
	| z \rangle = \frac{1}{\sqrt{1 + {\bar z} z}} ( | 0 \rangle + z | 1 \rangle ) 
	\label{coherentstatefor1/2} 
	\end{eqnarray}
stands for a single qubit in the Majorana representation \cite{Majorana} (the vector $| z \rangle$ is nothing but a SU(2) coherent state for a spin $j = \frac{1}{2}$ as can be seen by identifying the qubits $| n \rangle = | 0 \rangle$ and $| 1 \rangle$ to the spin states  
$| j , m ) = | \frac{1}{2} , \frac{1}{2})$ and $| \frac{1}{2} , -\frac{1}{2})$, respectively). By iteration of Eq.~(\ref{psiNpsiN-1}), we obtain
	\begin{eqnarray*}
| \psi_d \rangle = | z \rangle \otimes | z \rangle \otimes \cdots \otimes | z \rangle,   
	\end{eqnarray*}
with $N$ factors $| z \rangle$. 
\\

As a r\'esum\'e, we have the following result. If the qudit state $| \psi_d \rangle$ given by Eq.~(\ref{psiN}) is separable, then it can be written 
	\begin{eqnarray}
| \psi_d \rangle = \frac{1}{(1 + {\bar z} z)^{\frac{N}{2}}} 
\sum_{k=0}^N z^k \sqrt{\frac{N!}{k! (N-k)!}} | N ; k \rangle
                 = | z \rangle \otimes | z \rangle \otimes \cdots \otimes | z \rangle,   
	\label{productofNz}
	\end{eqnarray}
so that $| \psi_d \rangle$ is completely separable into the tensor product of $N$ identical SU(2) coherent states for a spin $j = \frac{1}{2}$. 

\subsection{Separable states and coherent states}

Let us consider the unitary displacement operator
	\begin{eqnarray*}
	D(\xi) = \exp( \xi q_i^+ - \bar\xi q_i^-), \quad \xi \in \mathbb{C}
	\end{eqnarray*}
for the $i$-th qubit. The action of $D(\xi)$ on the $i$-th qubit $| 0 \rangle$ can be calculated to be 
	\begin{eqnarray}
	D(\xi) | 0 \rangle = \cos (|\xi|) | 0 \rangle + \frac{\xi}{|\xi|} \sin (|\xi|) | 1 \rangle. 
	\label{bibi}
	\end{eqnarray}
By introducing
	\begin{eqnarray*}
	z = \frac{\xi}{|\xi|} \tan (|\xi|) \quad \Rightarrow \quad \cos^2 (|\xi|) = \frac{1}{1 + {\bar z} z}
	\end{eqnarray*}
in (\ref{bibi}), we obtain
	\begin{eqnarray*}
	D(\xi) | 0 \rangle = \frac{1}{\sqrt{1 + {\bar z} z}} ( | 0 \rangle + z | 1 \rangle ). 
	\end{eqnarray*}
Hence, we have 
\begin{eqnarray*}
D(\xi) | 0 \rangle = | z \rangle, 
\end{eqnarray*}
where $| z \rangle$ is the coherent state defined in (\ref{coherentstatefor1/2}). This well-known result can be extended to the case of $N$ qubits. The action of the operator $\exp( \xi q^+ - \bar\xi q^-)$, where $q^+$ and $q^-$ are given in (\ref{green}), on the Dicke state $|N ; 0 \rangle = |00\cdots0\rangle$ reads  
	\begin{eqnarray*}
	\exp( \xi q^+ - \bar\xi q^-) |N ; 0 \rangle = 
	| z \rangle \otimes | z \rangle \otimes \cdots \otimes | z \rangle.  
	\end{eqnarray*}
Therefore, the separable state $| \psi_d \rangle$ given by (\ref{productofNz}) can be written in three different forms, namely 
	\begin{eqnarray*}
	| \psi_d \rangle = \frac{1}{(1 + {\bar z} z)^{\frac{N}{2}}} \sum_{k=0}^N z^k \sqrt{\frac{N!}{k! (N-k)!}} | N ; k \rangle 
									 = | z \rangle \otimes | z \rangle \otimes \cdots \otimes | z \rangle 
	                 = \exp( \xi q^+ - \bar\xi q^-) | N ; 0 \rangle, 
	\end{eqnarray*}
where the last member coincides, modulo some changes of notation, with the Perelomov coherent state derived in \cite{DK} (see formulas (122) and (123) in \cite{DK}). 
\\

\section{Majorana description}

We now go back to the general case where the qudit state $| \psi_d \rangle$ of ${\cal G}_{d}$ is not necessarily a separable state. This state (normalized to unity) can be written in two different forms, namely, as in Eq.~(\ref{psiN})
	\begin{eqnarray}
	| \psi_d \rangle = c_0 | N ; 0 \rangle + c_1 | N ; 1 \rangle + \cdots + c_N | N ; N \rangle, \quad 
	N = d-1, \quad 
	\sum_{k=0}^N | c_k |^2 = 1  
	\label{formck}
	\end{eqnarray}
or, according to the Majorana description \cite{Majorana}, as 
	\begin{eqnarray}
	| \psi_d \rangle = {\cal N}_d \sum_{\sigma \in S_N} \sigma 
	( | z_1 \rangle \otimes | z_2 \rangle \otimes \cdots \otimes | z_N \rangle )
	\label{formcoherentstateN}
	\end{eqnarray}
(see Annexe for a discussion of the equivalence between (\ref{formck}) and (\ref{formcoherentstateN}) in the framework of the Bargmann function associated with 
$| \psi_d \rangle$ and the so-called Majorana stars). In Eq.~(\ref{formcoherentstateN}), the state $\vert z_i \rangle$ (with $i = 1,2, \cdots, N$) is given by 
(\ref{coherentstatefor1/2}) with $z = z_i$. Furthermore, ${\cal N}_d$ is a normalization factor and the sum over $\sigma$ runs here over all the permutations of the symmetric group $S_N$. The coefficients $c_0, c_1, c_2, \cdots, c_N$ can be expressed in terms of the coefficients ${\cal N}_d, z_1, z_2, \cdots, z_N$. The case where $N$ is arbitrary is rather intricate. Therefore, for pedagogical reasons we start with the case of $N = 2$ qubits.  

\subsection{The case $N = 2$}

For $N=2$ ($\Leftrightarrow d = 3$), on the one hand we have
	\begin{eqnarray*}
	| \psi_3 \rangle = c_0 | 2 ; 0 \rangle + c_1 | 2 ; 1 \rangle + c_2 | 2 ; 2 \rangle, \quad 
	|c_0|^2 + |c_1|^2 + |c_2|^2 = 1, 
	\end{eqnarray*}
where the Dicke states $| 2 ; k \rangle$ with $k = 0, 1, 2$ are 
	\begin{eqnarray}
	| 2 ; 0 \rangle  = | 00 \rangle, \quad
	| 2 ; 1 \rangle  = \frac{1}{\sqrt{2}} (| 01 \rangle + | 10 \rangle), \quad
	| 2 ; 2 \rangle  = | 11 \rangle.     
	\label{DickestatesforN=2}
	\end{eqnarray}
On the other hand 
	\begin{eqnarray*}
| \psi_3 \rangle = {\cal N}_3 ( | z_1 \rangle \otimes | z_2 \rangle  + 
                                | z_2 \rangle \otimes | z_1 \rangle  ). 
	\end{eqnarray*}
Therefore, we have to compare 
	\begin{eqnarray*}
	| \psi_3 \rangle = 
	c_0 | 00 \rangle + c_1 \frac{1}{\sqrt{2}} (| 01 \rangle + | 10 \rangle) + c_2 | 11 \rangle       
	\end{eqnarray*}
with 	
	\begin{eqnarray*}
	| \psi_3 \rangle = {\cal N}_3 \frac{1}{\sqrt{1 + |z_1|^2}} \frac{1}{\sqrt{1 + |z_2|^2}}  
	\left[ 2 | 00 \rangle + (z_1 + z_2) (| 01 \rangle + | 10 \rangle) + 2 z_1 z_2  | 11 \rangle \right]. 
	\end{eqnarray*}
This leads to
	\begin{eqnarray}
	\frac{1}{2}        c_0 &=& 
	{\cal N}_3 \frac{1}{\sqrt{1 + |z_1|^2}} \frac{1}{\sqrt{1 + |z_2|^2}},                        \nonumber \\
	\frac{1}{\sqrt{2}} c_1 &=& 
	{\cal N}_3 \frac{1}{\sqrt{1 + |z_1|^2}} \frac{1}{\sqrt{1 + |z_2|^2}}  (z_1 + z_2),   \label{c0c1c2}    \\
	\frac{1}{2}        c_2 &=& 
	{\cal N}_3 \frac{1}{\sqrt{1 + |z_1|^2}} \frac{1}{\sqrt{1 + |z_2|^2}}  z_1 z_2.               \nonumber
	\end{eqnarray}
Of course, the complex numbers $z_1$ and $z_2$ are the roots of the equation 
	\begin{eqnarray}
	z^2 - (z_1 + z_2) z + z_1 z_2 = 0. 
	\label{eqndegree2}
	\end{eqnarray}
Therefore, by combining Eqs.~(\ref{c0c1c2}) and (\ref{eqndegree2}), we end up with the quadratic equation 
	\begin{eqnarray}
	c_0 z^2 - \sqrt{2} c_1 z + c_2 = 0, 
	\label{eqndegree2enc0c1c2}
	\end{eqnarray}
so that $z_1$ and $z_2$ are given by 
	\begin{eqnarray}
	z_1 = z_+, \quad z_2 = z_-, \quad z_{\pm} = \frac{c_1 \pm \sqrt{c_1^2 - 2 c_0 c_2}}{\sqrt{2} c_0}   
	\label{deuxracinespourN=2}
	\end{eqnarray}
for $c_0 \not=0$ ($z = \frac{1}{\sqrt{2}} \frac{c_2}{c_1}$ for $c_0 = 0$). Observe that, when the so-called concurrence $C$ defined by (see Ref.~\cite{Concurrence})
	\begin{eqnarray}
	C = | c_1^2 - 2 c_0 c_2 |   
	\label{concurrenceforN=2}
	\end{eqnarray}
vanishes, we have $z_1 = z_2 = z$. Therefore, the state 
	\begin{eqnarray*}
| \psi_3 \rangle = | z \rangle \otimes | z \rangle = 
\frac{1}{1 + {\bar z} z} \left[ | 00 \rangle + z (| 01	 \rangle + | 10 \rangle) + z^2  | 11 \rangle \right] 
	\end{eqnarray*}
is separable. 

\subsection{The case $N$ arbitrary} \label{general case}

The case $N$ arbitrary is very much involved. Equations (\ref{c0c1c2}) and (\ref{eqndegree2enc0c1c2}) for $N = 2$ can be generalized as follows. In the general case of $N$ qubits, the vector $| \psi_d \rangle$ of the space ${\cal G}_{d}$, normalized via $\langle \psi_d | \psi_d \rangle = 1$, is given by (\ref{formck}) in terms of Dicke states or by (\ref{formcoherentstateN}) in the Majorana representation. The coefficients $c_0, c_1, c_2, \cdots, c_N$ are connected to the complex numbers 
${\cal N}_d, z_1, z_2, \cdots, z_N$ through 
	\begin{eqnarray*}
	c_k = N! N_1 N_2 \cdots N_N {\cal N}_d \sqrt{\frac{k! (N-k)!}{N!}} s_k(z_1 z_2 \cdots z_N), 
	\end{eqnarray*}
where $s_k(z_1 z_2 \cdots z_N)$ is the elementary symmetric polynomial (invariant under $S_N$) in $N$ variables $z_1,z_2, \cdots, z_N$ defined as 
	\begin{eqnarray*} 
	s_0(z_1 z_2 \cdots z_N) = 1, \quad 
	s_k(z_1 z_2 \cdots z_N) = \sum_{1 \leq i_1 < i_2 < \cdots < i_k \leq N} z_{i_1} z_{i_2} \cdots z_{i_k}, 
	\quad k = 1, 2, \cdots, N
	\end{eqnarray*}
and the normalization factors $N_1, N_2, \cdots, N_N, {\cal N}_d$ are given by 
	\begin{eqnarray} 
N_i = \frac{1}{\sqrt{1 + \bar z_{i} z_i}}, \quad i = 1, 2, \cdots, N  
  \label{1normalizationcalNindiceN}
	\end{eqnarray}	
and 
	\begin{eqnarray}
| {\cal N}_d |^{-2}	= N! \sum_{\sigma \in S_N} \prod_{i = 1}^N \langle z_i | z_{\sigma(i)} \rangle, \quad 
\langle z_i | z_{\sigma(i)} \rangle = \frac{ 1 + \bar z_{i} z_{\sigma(i)} } { \sqrt{(1 + \bar z_{i} z_i)(1 + \bar z_{{\sigma(i)}} z_{\sigma(i)})} }. 
\label{2normalizationcalNindiceN}
	\end{eqnarray}
Note that 
	\begin{eqnarray}
| {\cal N}_d |^{-2}	= N! {\rm perm} (A_N),
	\label{norme-et-per}
	\end{eqnarray}
where 
	\begin{eqnarray}
{\rm perm} (A_N) = \sum_{\sigma \in S_N} \prod_{i = 1}^N \langle z_i | z_{\sigma(i)} \rangle 
                 = \frac{1}{\prod_{j = 1}^N (1 + {\bar z_j} z_j)} \sum_{\sigma \in S_N} \prod_{i = 1}^N ( 1 + \bar z_{i} z_{\sigma(i)} )
\label{defpermanent}
	\end{eqnarray}
stands for the permanent of the $N \times N$ matrix $A_N$ of elements 
	\begin{eqnarray*}
(A_N)_{ij} = \langle z_i | z_j \rangle, \quad i, j = 1, 2, \cdots, N. 
	\end{eqnarray*}
Finally, for fixed $c_0, c_1, \cdots, c_N$, the numbers $z_1, z_2, \cdots, z_N$ are the roots 
(Majorana roots) of the polynomial equation of degree $N$ 
	\begin{eqnarray}
\sum_{k=0}^N (-1)^k \sqrt{\frac{N!}{k! (N-k)!}} c_k z^{N-k} = 0,   
	\label{d'Alembert}
	\end{eqnarray}
which generalizes (\ref{eqndegree2enc0c1c2}).
\\

The complete proof of (\ref{d'Alembert}) is based on the fact that two generic qubit states 
$\vert z_i \rangle$ and 
$\vert z_j \rangle$, with $j \not= i$, 
are orthogonal if and only if the variables $z_i$ and $z_j$ satisfy  $z_j = - \frac{1}{\bar z_{i}}$. The state 
$\vert \psi_d \rangle$ is orthogonal to the 
$N$ states $\vert - \frac{1}{\bar z_{i}} \rangle \otimes  
            \vert - \frac{1}{\bar z_{i}} \rangle \otimes \cdots \otimes 
						\vert	- \frac{1}{\bar z_{i}} \rangle$
for $i = 1, 2, \cdots, N$. This orthogonality condition shows that the variables $z_i$ are indeed solutions of Eq.~(\ref{d'Alembert}). 
\\

To sum up, we have the following central result. Any vector $| \psi_d \rangle$ in the space ${\cal G}_d$ reads
	\begin{eqnarray}
| \psi_d \rangle = \sum_{k = 0}^N c_k | N ; k \rangle \ \Leftrightarrow \ 
| \psi_d \rangle = N! N_1 N_2 \cdots N_N {\cal N}_d \sum_{k = 0}^N \sqrt{\frac{k! (N-k)!}{N!}} s_k(z_1 z_2 \cdots z_N) | N ; k \rangle, 
	\label{psienSchur}
	\end{eqnarray}
where the normalization factors $N_1, N_2, \cdots, N_N$, and ${\cal N}_d$ (with $d = N+1$) can be calculated from Eqs.~(\ref{1normalizationcalNindiceN}) and 
(\ref{2normalizationcalNindiceN}) and the variables $z_1, z_2, \cdots, z_N$ are given in terms of $c_0, c_1, \cdots, c_N$ by Eq.~(\ref{d'Alembert}). Note that 
(\ref{2normalizationcalNindiceN}) can be rewritten as 
	\begin{eqnarray*}
| {\cal N}_d |^{-2}	= \frac{N!}{(1 + \bar z_{1} z_1)
                                (1 + \bar z_{2} z_2) \cdots 
																(1 + \bar z_{N} z_N)} \sum_{\sigma \in S_N} \prod_{i = 1}^N ( 1 + \bar z_{i} z_{\sigma(i)} ), 
	\end{eqnarray*}
so that 
	\begin{eqnarray*}
| N! N_1 N_2 \cdots N_N {\cal N}_d |^{-2} = \frac{1}{N!} \sum_{\sigma \in S_N} \prod_{i = 1}^N ( 1 + \bar z_{i} z_{\sigma(i)} ).
	\end{eqnarray*}
Therefore, Eq.~(\ref{psienSchur}) becomes
	\begin{eqnarray}
| \psi_d \rangle = \sqrt{\frac{N!}{\sum_{\sigma \in S_N} \prod_{i = 1}^N ( 1 + \bar z_{i} z_{\sigma(i)} )}} 
\sum_{k = 0}^N \sqrt{\frac{k! (N-k)!}{N!}} s_k(z_1 z_2 \cdots z_N) | N ; k \rangle
	\label{psienSchurbis}
	\end{eqnarray}
up to a phase factor. 
\\
      
As a check of the last result, note that the introduction of (\ref{ck}) into (\ref{d'Alembert}) yields a trivial identity. Furthermore, in the particular case where the solutions of (\ref{d'Alembert}) are identical, i.e., 
	\begin{eqnarray*}
z = z_1 = z_2 = \cdots = z_N \ \Rightarrow \ s_k(z z \cdots z) = \frac{N!}{k! (N-k)!} z^k,
	\end{eqnarray*}
then Eq.~(\ref{psienSchurbis}) leads to the completely separable state (\ref{productofNz}). In this particular case, from Eq.~(\ref{defpermanent}) we have 
	\begin{eqnarray*}
{\rm perm} (A_N) = N!
	\end{eqnarray*}
(which is the maximum value of ${\rm perm} (A_N)$). Therefore, in the general case the quantity 
	\begin{eqnarray}
P_d = \frac{1}{N!} {\rm perm} (A_N) = \frac{1}{N!} \sum_{\sigma \in S_N} \prod_{i = 1}^N \langle z_i | z_{\sigma(i)} \rangle
	\label{def-Pd}
	\end{eqnarray}
can be used for characterizing the degree of entanglement of the state $| \psi_d \rangle$.  

\subsection{The cases $d = 2$, $3$, $4$, and $5$}

\subsubsection{Case $d = 2$} 
The state
	\begin{eqnarray*}
| \psi_2 \rangle = c_0 | 1 ; 0 \rangle + c_1 | 1 ; 1 \rangle, \quad 
| 1 ; 0 \rangle = | 0 \rangle, \quad | 1 ; 1 \rangle = | 1 \rangle
	\end{eqnarray*}
is the most general qubit (linear combination of the basic qubits $| 0 \rangle$ and $| 1 \rangle$). Of course, the notion of separability does not apply in this case. 

\subsubsection{Case $d = 3$} 

The general normalized qutrit vector is 
	\begin{eqnarray*}
| \psi_3 \rangle = 
c_0 | 2 ; 0 \rangle + c_1 | 2 ; 1 \rangle + c_2 | 2 ; 2 \rangle, \quad \sum_{k=0}^2 |c_k|^2 = 1, 
	\end{eqnarray*}
where the Dicke states $| 2 ; k \rangle$ with $k = 0,1,2$ are 
\begin{eqnarray*}
	| 2 ; 0 \rangle  = | 0 \rangle \otimes | 0 \rangle, \quad
	| 2 ; 1 \rangle  = 
	\frac{1}{\sqrt{2}} ( | 0 \rangle \otimes | 1 \rangle + | 1 \rangle \otimes | 0 \rangle ), \quad
	| 2 ; 2 \rangle  = | 1 \rangle \otimes | 1 \rangle,     
	\end{eqnarray*}
cf.~(\ref{DickestatesforN=2}). In the Majorana description, Eqs.~(\ref{formcoherentstateN}), (\ref{norme-et-per}), and (\ref{def-Pd}) gives
  \begin{eqnarray*}
	| \psi_3 \rangle = 
	\frac{1}{2 \sqrt{P_3}} ( | z_1 \rangle \otimes | z_2 \rangle + | z_2 \rangle \otimes | z_1 \rangle ), 
	\end{eqnarray*}	
with 
	\begin{eqnarray*}
P_3 &=& \frac{1}{2} {\rm perm} (A_2), \\ 
    &=& \frac{1}{2} ( 1 + | \langle z_1 | z_2 \rangle |^2 ), \\ 
		&=& \frac{1}{2} 
		\frac{(1 + \bar z_{1}{z_1})(1 + \bar z_{2}{z_2}) + (1 + \bar z_{1}{z_2})(1 + \bar z_{2}{z_1})} 
		     {(1 + \bar z_{1}{z_1})(1 + \bar z_{2}{z_2})},
 	\end{eqnarray*}
where $z_1$ and $z_2$ are the roots (\ref{deuxracinespourN=2}) of the quadratic equation (\ref{eqndegree2enc0c1c2}). 
\\

It can be shown that 
	\begin{eqnarray*}
\vert \langle z_1 \vert z_2 \rangle \vert^2 = \frac{1-C}{1+C} \ \Leftrightarrow \ 
P_3 = \frac{1}{1 + C} \ \Leftrightarrow \ C = \frac{1}{P_3} - 1, 
 	\end{eqnarray*}
where the concurrence $C$ for a two-qubit system is defined by Eq.~(\ref{concurrenceforN=2}). Thus, another expression for $C$ is 
	\begin{eqnarray*}
C = \frac{(1 + \bar z_{1}{z_1})(1 + \bar z_{2}{z_2}) - (1 + \bar z_{1}{z_2})(1 + \bar z_{2}{z_1})} 
		     {(1 + \bar z_{1}{z_1})(1 + \bar z_{2}{z_2}) + (1 + \bar z_{1}{z_2})(1 + \bar z_{2}{z_1})}.
 	\end{eqnarray*}
The possible values of $C$ and $P_3$ are 
	\begin{eqnarray*}
\frac{1}{2} \leq P_3 \leq 1 \ \Leftrightarrow \ 1 \geq C \geq 0. 
 	\end{eqnarray*}
Therefore, a vanishing concurrence $C = 0$ (which reflects the absence of entanglement) corresponds 
to $P_3 = 1$; in the particular case $P_3 = 1 \Leftrightarrow C = 0$, we have 
$z = z_1 = z_2 \ \Leftrightarrow \ \langle z_1 | z_2 \rangle = 1$ that leads to the separable state 
$| \psi_3 \rangle = | z \rangle \otimes | z \rangle$. Furthermore, for $C=1$ (which characterizes entangled states), we have $P_3 = \frac{1}{2} \ \Leftrightarrow \ \langle z_1 | z_2 \rangle = 0$. Consequently, in the general case ($z_1$ and $z_2$ arbitrary), $P_3$ constitutes an alternative to the concurrence $C$ for measuring the degree of entanglement of the general qutrit $| \psi_3 \rangle$.
\\

It is interesting to note that $P_3$ can be alternatively written as 
$$ 
P_3 = \frac{1}{4} (3 + n_1.n_2)
$$ 
where the vectors 
\begin{equation}
n_k = \bigg(\frac{ z_k +\bar z_k}{1+ z_k\bar z_k}, -{\rm i} \frac{ z_k -\bar z_k}{1+ z_k\bar z_k}, \frac{ 1-z_k \bar z_k}{1+ z_k\bar z_k}\bigg)
\label{vect-n}
\end{equation}
(with $k = 1, 2$ and ${\rm i} = \sqrt{-1}$) are unit vectors in the space $\mathbb{R}^3$ which serve to locate points on the Bloch sphere. Therefore, entangled states are obtained for $n_1.n_2 = -1$ (in this case, $P_3$ takes its minimal value $\frac{1}{2}$).
\\

Note the following relation 
\begin{equation}
 n_i.n_j = 2 \vert \langle z_i \vert z_j \rangle\vert^2 - 1
\label{prod-vect-n}
\end{equation}
valid for arbitrary $i$ and $j$. This relation will be useful for deriving closed-form expressions of $P_d$ in higher dimensional cases. 

\subsubsection{Case $d = 4$} 

In this case, the general state $| \psi_4 \rangle$ of ${\cal G}_4$ is made of $N=3$ qubits. It takes the form
	\begin{eqnarray*}
| \psi_4 \rangle = c_0 | 3 ; 0 \rangle + c_1 | 3 ; 1 \rangle + c_2 | 3 ; 2 \rangle + c_3 | 3 ; 3 \rangle, \quad \sum_{k=0}^3 |c_k|^2 = 1,
 	\end{eqnarray*}
where the Dicke states $| 3 ; k \rangle$ with $k = 0,1,2,3$ are 
\begin{eqnarray*}
	| 3 ; 0 \rangle  &=& | 0 \rangle \otimes | 0 \rangle \otimes | 0 \rangle, \\
	| 3 ; 1 \rangle  &=& \frac{1}{\sqrt{3}} (| 0 \rangle \otimes | 0 \rangle \otimes | 1 \rangle + 
	                                         | 0 \rangle \otimes | 1 \rangle \otimes | 0 \rangle +
																				   | 1 \rangle \otimes | 0 \rangle \otimes | 0 \rangle  ), \\
	| 3 ; 2 \rangle  &=& \frac{1}{\sqrt{3}} (| 0 \rangle \otimes | 1 \rangle \otimes | 1 \rangle + 
	                                         | 1 \rangle \otimes | 0 \rangle \otimes | 1 \rangle +
																				   | 1 \rangle \otimes | 1 \rangle \otimes | 0 \rangle  ), \\
	| 3 ; 3 \rangle  &=& | 1 \rangle \otimes | 1 \rangle \otimes | 1 \rangle. 
	\end{eqnarray*}
In the Majorana representation, we have 
	\begin{eqnarray*}
|{\cal N}_4|^{-1} | \psi_4 \rangle 
&=& | z_1 \rangle \otimes | z_2 \rangle \otimes | z_3 \rangle + 
    | z_2 \rangle \otimes | z_1 \rangle \otimes | z_3 \rangle + 
	  | z_1 \rangle \otimes | z_3 \rangle \otimes | z_2 \rangle   \\						
&+& | z_3 \rangle \otimes | z_2 \rangle \otimes | z_1 \rangle + 
    | z_2 \rangle \otimes | z_3 \rangle \otimes | z_1 \rangle + 
		| z_3 \rangle \otimes | z_1 \rangle \otimes | z_2 \rangle,                                
 	\end{eqnarray*}
where the states $| z_i \rangle$ are given by (\ref{coherentstatefor1/2}) with $z = z_1, z_2, z_3$ and the complex numbers $z_1, z_2, z_3$ are solutions of the polynomial equation of degree 3 
	\begin{eqnarray*}
c_0 z^3 - \sqrt{3} c_1 z^2 + \sqrt{3} c_2 z - c_3 = 0. 
 	\end{eqnarray*}
The normalization factor ${\cal N}_4$ reads
	\begin{eqnarray*}
|{\cal N}_4|^{-2} = 3! {\rm perm}(A_3) = (3!)^2 P_4,  
 	\end{eqnarray*} 
with
		\begin{eqnarray*}
P_4 = \frac{1}{6} \left( 1 + \vert \langle z_1 \vert z_2 \rangle \vert^2 + 
                             \vert \langle z_2 \vert z_3 \rangle \vert^2 + 
														 \vert \langle z_3 \vert z_1 \rangle \vert^2 +
\langle z_1 \vert z_2 \rangle \langle z_2 \vert z_3 \rangle \langle z_3 \vert z_1 \rangle + 
\langle z_1 \vert z_3 \rangle \langle z_3 \vert z_2 \rangle \langle z_2 \vert z_1 \rangle \right)
		\end{eqnarray*}
or alternatively
$$ 
P_4 = \frac{1}{6} (3 + n_1.n_2 + n_2.n_3 + n_3.n_1),
$$ 
where the components of the vectors $n_i$ ($i= 1, 2, 3$) are given by (\ref{vect-n}). From 
Eq.~(\ref{prod-vect-n}), we get 
$$
P_4 = \frac{1}{3} (\vert \langle z_1 \vert z_2 \rangle\vert^2 + 
                   \vert \langle z_2 \vert z_3 \rangle\vert^2 + 
									 \vert \langle z_3 \vert z_1 \rangle\vert^2)
$$
that clearly shows that 
	\begin{eqnarray*}
\frac{1}{3} \leq P_4 \leq 1
 	\end{eqnarray*}
The case of complete separability corresponds to $P_4 = 1$. The minimal value $P_4 = \frac{1}{3}$ is obtained for entangled states. 

\subsubsection{Case $d = 5$} 

In this case, the variables $z_i$ ($i = 1, 2, 3, 4$) are solutions of the equation of degree 4 
$$ 
c_0 z^4 - \sqrt{4} c_1 z^3 + \sqrt{6} c_2 z^2 - \sqrt{4} c_3 z + c_4 = 0.
$$
The calculation of $P_5$ yields
	\begin{eqnarray*}
P_5 &=& \frac{1}{4!} \bigg( - 6 + 4(\vert \langle z_1 \vert z_2 \rangle\vert^2  + 
         \vert \langle z_1 \vert z_3 \rangle\vert^2  +
         \vert \langle z_1 \vert z_4 \rangle\vert^2  +
         \vert \langle z_2 \vert z_3 \rangle\vert^2  + 
	       \vert \langle z_2 \vert z_4 \rangle\vert^2  +
         \vert \langle z_3 \vert z_4 \rangle\vert^2)  \\
    &+&2(\vert \langle z_1 \vert z_2 \rangle\vert^2 \vert \langle z_3 \vert z_4 \rangle\vert^2 + 
	       \vert \langle z_1 \vert z_3 \rangle\vert^2 \vert \langle z_2 \vert z_4 \rangle\vert^2 +
         \vert \langle z_1 \vert z_4 \rangle\vert^2 \vert \langle z_2 \vert z_3 \rangle\vert^2) \bigg)
	\end{eqnarray*}	
or	
	\begin{eqnarray*}
P_5 &=& \frac{1}{4!} \bigg( \frac{15}{2} + \frac{5}{2} \big( n_1.n_2 + n_1.n_3 + n_1.n_4 + n_2.n_3 + n_2.n_4 + n_3.n_4    \big) \\
    &+& \frac{1}{2}                                    \big( (n_1.n_2)(n_3.n_4) + (n_1.n_3)(n_2.n_4) + (n_1.n_4)(n_2.n_3) \big) \bigg)
	\end{eqnarray*}
with
$$
\frac{1}{4} \leq P_5 \leq 1. 
$$
The minimal value $P_5 = \frac{1}{4}$ can be obtained from 
	\begin{eqnarray}
 \langle z_1 \vert z_2 \rangle = 
 \langle z_1 \vert z_3 \rangle = 
 \langle z_1 \vert z_4 \rangle = 0 
	\label{equality}
	\end{eqnarray}
or from any analogue equality deduced from (\ref{equality}) by permutations of the indices $1, 2, 3, 4$.

\subsubsection{Case $d$ arbitrary}

The general case is approached in Section (\ref{general case}). For $d$ arbitrary, it can be shown that 
	\begin{eqnarray*}
\frac{1}{N} \leq  P_d  \leq 1, 
 	\end{eqnarray*}
the situation where $P_d = 1$ corresponding to complete separability and $P_d = \frac{1}{N}$ to entangled states. Therefore, the parameter $P_d$ can serve as a measure of the entanglement of the symmetric qudit state $| \psi_d \rangle$ described by $N = d-1$ qubits.
\\

The minimal value of $P_d$ can be obtained when 
\begin{equation}
\langle z_1\vert z_2 \rangle = \langle z_1\vert z_3 \rangle = \cdots = \langle z_1\vert z_N \rangle = 0.
\label{cond-N}
\end{equation}
Thus, Eq.~(\ref{def-Pd}) can be reduced to 
$$
P_d = \frac{1}{N!} \sum_{\sigma \in S_{N-1}} \prod_{i=2}^{N}\langle z_i\vert z_{\sigma(i)} \rangle.
$$
The condition (\ref{cond-N}) implies that 
$$
z_2 = z_3 = \cdots =z_N.
$$
In this case, we have
$$
\sum_{\sigma \in S_{N-1}} \prod_{i=2}^{N}\langle z_i\vert z_{\sigma(i)} \rangle = (N-1)!
$$
and the minimal value of $P_d$ is 
$$ 
P_d = \frac{(N-1)!}{N!} = \frac{1}{N}. 
$$
The same result can be obtained equally well, due to the invariance of $P_d$ under permutation symmetry, from any of the following conditions
\begin{eqnarray*}
&&\langle z_2\vert z_1 \rangle = \langle z_2\vert z_3 \rangle = \cdots = \langle z_2\vert z_N \rangle = 0, \\
&&\langle z_3\vert z_1 \rangle = \langle z_3\vert z_2 \rangle = \cdots = \langle z_3\vert z_N \rangle = 0, \\
&&\vdots \\
&&\langle z_N\vert z_1 \rangle = \langle z_N\vert z_2 \rangle = \cdots = \langle z_N\vert z_{N-1} \rangle = 0.
\end{eqnarray*}
instead of the condition (\ref{cond-N}). 

\section{Fubini-Study metric}

\subsection{The separable case}

The adequate approach to deal with the geometrical properties of a quantum state manifold is based on the derivation of the corresponding Fubini-Study metric 
\cite{Fubini-Study}. The Fubini-Study metric is defined by the infinitesimal distance $ds$ between two neighboring quantum states. This derivation is simplified by adopting the coherent states formalism. Indeed, for a single qubit coherent state $\vert z \rangle$ this is realized in the following way. Let us define the K\"ahler potential 
$K(\bar z ; z)$ as
	\begin{eqnarray}
	K(\bar z ; z) = \ln ( \langle 0 \vert z \rangle )^{-2}. 
	\label{K for state with 1 qubit}
 	\end{eqnarray}
Using the expression (\ref{coherentstatefor1/2}) of the coherent state $\vert z \rangle$, we have
		\begin{eqnarray*}
K(\bar z ; z) = \ln ( 1 + {\bar z} z)
		\end{eqnarray*}
and the metric tensor
$$ 
g = \frac{{\partial}^2 K}{\partial z \partial \bar z}
$$
becomes
$$ 
g = \frac{1}{( 1 + {\bar z} z)^2}, 
$$
so that the Fubini-Study metric $ds^2$ reads
$$ 
ds^2 = g dz d{\bar z} = \frac{1}{( 1 + {\bar z} z)^2} dz d{\bar z},
$$
which coincides with the metric of the unit sphere. This provides us with a simple way to describe the $2$-sphere $S^2$, or equivalently the complex projective space $CP^1$, usually regarded as the space of states of a $\frac{1}{2}$-spin particle. 
\\

This can be generalized to the completely separable state 
$$ 
\vert z_1 z_2 \cdots z_N \rangle = \vert z_1 \rangle \otimes \vert z_2 \rangle \otimes \cdots \otimes \vert z_N \rangle
$$
constructed from the tensor product of $N$ qubit coherent states. In this case, the K\"ahler potential is given by
		\begin{eqnarray}
K(\bar z_{1} \bar z_{2} \cdots \bar z_{N} ; z_1 z_2 \cdots z_N) = 
\ln ( \langle 0 0 \cdots 0 \vert z_1 z_2 \cdots z_N \rangle )^{-2}. 
		\label{K for separable state with N qubits} 
		\end{eqnarray}
This leads to 
		\begin{eqnarray}
K(\bar z_{1} \bar z_{2} \cdots \bar z_{N} ; z_1 z_2 \cdots z_N) = \sum_{i = 1}^N \ln ( 1 + {\bar z_i} z_i).  
		\label{K for separable state with N qubits-bis}
		\end{eqnarray}
The metric tensor $g$ is defined via its components 
$$
g^{ij} = \frac{{\partial}^2 K}{\partial z_i \partial \bar z_j} \ \Rightarrow \ g^{ij} = \delta_{i,j} \frac{1}{(1 + \bar z_i z_i)^2}. 
$$
Finally, the Fubini-Study line element $ds^2$ is
		\begin{equation}
ds^2 = g^{ij} dz_i d\bar z_j  = \sum_{i=1}^{N} \frac{1}{( 1 + \bar z_i z_i)^2} dz_i d\bar z_i
		\label{metric-sep}
		\end{equation}
associated with the complex space $CP^1\times CP^1\times\cdots \times CP^1$.  
\\

In the special case where the complex variables $z_i$ are identical, i.e., $z_1 = z_2 = \cdots = z_N = z$, the state $\vert z_1 z_2 \cdots z_N \rangle$ reduces
to the coherent state given by (\ref{productofNz}). In this case, the Fubini-Study metric takes the form 
$$ 
ds^2 = N \frac{1}{( 1 + \bar z z)^2} dz d\bar z,
$$
which describes the unit $2$-sphere, of radius $\sqrt{N}$, written in stereographic coordinates.

\subsection{The arbitrary case}

We now apply the just described geometrical picture to calculate the Fubini-Study metric for an arbitrary multiqubit symmetric state $\vert \psi_d \rangle$. Here, we define the K\"ahler potential through 
		\begin{eqnarray*}
K(\bar z_{1} \bar z_{2} \cdots \bar z_{N} ; z_1 z_2 \cdots z_N) = \ln ( \langle 0 0 \cdots 0 \vert \psi_d \rangle )^{-2}
		\end{eqnarray*}
as a generalization of (\ref{K for state with 1 qubit}) and (\ref{K for separable state with N qubits}). It is easy to show that 
		\begin{eqnarray*}
( \langle 0 0 \cdots 0 \vert \psi_d \rangle )^{-2} 
&=& \frac{1}{N!} \sum_{\sigma \in S_N} \prod_{i = 1}^N \langle z_i | z _{\sigma(i)} \rangle 
\prod_{j = 1}^N ( 1 + \bar z_{j} z_{j} ), \\
&=& \frac{1}{N!} {\rm perm} (A_N)      \prod_{i = 1}^N ( 1 + \bar z_{i} z_{i} ), \\
&=& P_d \prod_{i = 1}^N ( 1 + \bar z_{i} z_{i} ). 
		\end{eqnarray*}                                               
Hence, we otain 
		\begin{eqnarray*}
K(\bar z_{1} \bar z_{2} \cdots \bar z_{N} ; z_1 z_2 \cdots z_N) = 
\ln P_d + \sum_{i = 1}^N \ln ( 1 + \bar z_{i} z_{i} )
		\end{eqnarray*} 
in terms of the parameter $P_d$ defined by (\ref{def-Pd}). As a result, the K\"ahler potential splits into two parts: one term is the K\"ahler potential corresponding to a completely separable state involving $N$ qubits, cf.~(\ref{K for separable state with N qubits-bis}), and the other term depends exclusively on the parameter $P_d$ which characterizes the degree of entanglement of the state $| \psi_d \rangle$. Then, the components of the corresponding metric tensor $g$ are 
		\begin{eqnarray*}
g^{ij} = \frac{{\partial}^2 K}       {\partial z_i \partial \bar z_j} 
       = \frac{{\partial}^2 \ln P_d} {\partial z_i \partial \bar z_j} + 
\delta_{i,j} \frac{1}{(1 + {\bar z}_i z_i)^2} 
		\end{eqnarray*}
and the Fubini-Study line element $ds^2$ is
		\begin{eqnarray*}
ds^2 = g^{ij} dz_i d\bar z_j  = 
\frac{{\partial}^2 \ln P_d} {\partial z_i \partial \bar z_j} dz_i d\bar z_j + 
\sum_{i=1}^{N} \frac{1}{( 1 + \bar z_i z_i)^2} dz_i d\bar z_i. 
		\end{eqnarray*}
In the special case where $P_d = 1$, corresponding to a completely separable state, the last equation gives back (\ref{metric-sep}) valid for a multiqubit separable state. This is a further indication that the parameter $P_d$ encodes the geometrical aspects due to the entanglement of a multiqubit symmetric state.

\section*{Annexe: Majorana stars and zeros of the Bargmann function}

\subsection*{The main idea}

An arbitrary normalized state $\vert \psi_d \rangle$ of the space ${\cal G}_d$ can be written either in terms of the Dicke states $|N ; k \rangle$ with $k = 0, 1, \cdots, N = d - 1$ (see Eq.~(\ref{formck})) or in terms of the coherent states $| z_i \rangle$ for $i = 1, 2, \cdots, N$ (see Eq.~(\ref{formcoherentstateN})). The variables $z_i$, called Majorana stars \cite{Majoranastars}, can be determined from the zeros of the Bargmann function 
$\psi : z \mapsto \psi(z)$ associated with the state $\vert \psi_d \rangle$. In fact, denoting by $\omega_i$ the zeros of the Bargmann function $\psi$, we shall show that the Majorana stars $z_i$ can be obtained from  the Bargmann zeros $\omega_i$ via  
		\begin{eqnarray*}
		z_i = - \frac{1}{\omega_i}, \quad i = 1, 2, \cdots, k_{\rm max} \leq N
		\end{eqnarray*} 
and we shall give the equation satisfied by the variables $z_i$.

\subsection*{Determining the Bargmann zeros}

In the analytic Fock-Bargmann representation \cite{Bargmannrepresentation}, an arbitrary normalized state 
$| \psi_d \rangle$ of ${\cal G}_d$ is represented by the Bargmann function $\psi$ defined by
\begin{equation}
\psi(z) = \langle N : \bar z \vert \psi_d \rangle,
\label{expression-psi1}
\end{equation}
where the bra $\langle N : \bar z|$ follows from the coherent state 
$$
\vert N : z \rangle = \frac{1}{(1+\bar z z)^{\frac{N}{2}}} \sum_{k=0}^{N} z^k \sqrt{\frac{N!}{k!(N-k)!}} \vert N ; k \rangle 
                    = | z \rangle \otimes | z \rangle \otimes \cdots \otimes | z \rangle 
$$
corresponding to the completely separable state (\ref{productofNz}). Thus, we have 
$$
\psi(z) = \frac{1}{(1+\bar z z)^{\frac{N}{2}}} \sum_{k=0}^{N} \sqrt{\frac{N!}{k!(N-k)!}} c_k z^k, 
$$
which can be decomposed as 
\begin{eqnarray}
\psi(z) = \frac{1}{(1+\bar z z)^{\frac{N}{2}}} {\cal P}(z),
\label{psi(z)enP(z)}
\end{eqnarray}
where
$$
{\cal P}(z) = \sum_{k=0}^{N} d_k z^k,  
$$
with
$$
d_k = \sqrt{\frac{N!}{k!(N-k)!}} c_k.
$$
In fact, the polynomial  
\begin{eqnarray}
{\cal P}(z) = \sum_{k=0}^{N} \sqrt{\frac{N!}{k!(N-k)!}} c_k z^k
\label{detailed expression of P(z)}
\end{eqnarray}
is of degree $k_{\rm max} \leq N$, where $k_{\rm max}$ is the maximum value of the index $k$ for which $c_k \not= 0$. Therefore, the polynomial ${\cal P}(z)$ can be factorized as
$$
{\cal P}(z) = d_{k_{\rm max}} (z - \omega_1)(z - \omega_2)\cdots (z - \omega_{k_{\rm max}}) 
$$
where $\omega_i$ ($i = 1, 2, \cdots, k_{\rm max}$) are called the Bargmann zeros. 

\subsection*{Expression of $\vert \psi_d \rangle$ in terms of the Bargmann zeros}

We now look for the expression of the state vector $\vert \psi_d \rangle$ in terms of  the Bargmann zeros 
$\omega_i$ ($i = 1, 2, \cdots, k_{\rm max}$). To this end, we remark that the scalar product between the state 
$$
\vert \omega_i \rangle 
= \frac{1}{\sqrt{1 + {\bar \omega_i} \omega_i}} (\vert 1\rangle  - \omega_i \vert 0 \rangle)
$$
and the coherent state $\vert \bar z \rangle \equiv \vert 1 : \bar z \rangle$ (see Eq.~(\ref{coherentstatefor1/2})) is
$$
\langle {\bar z} \vert \omega_i \rangle = \frac{z -  \omega_i}{\sqrt{(1 + {\bar z} z )(1 + {\bar \omega_i} \omega_i)}}, \quad i = 1, 2, \cdots, k_{\rm max}. 
$$
Thus, the polynomial ${\cal P}(z)$ can be written as
$$
{\cal P}(z) = 
d_{k_{\rm max}} (1 + \bar z z)^{\frac{k_{\rm max}}{2}}\prod_{i=1}^{k_{\rm max}} \sqrt{1 + {\bar \omega_i} \omega_i} \langle {\bar z} \vert \omega_i \rangle.
$$
Furthermore, by noting that 
$$
\langle {\bar z} \vert  0 \rangle = \frac{1}{\sqrt{1 + {\bar z} z}},
$$
we extend the definition of the states $\vert \omega_i \rangle$ (initially defined for 
$i = 1, 2, \cdots, k_{\rm max}$) by taking 
$$
\vert \omega_i \rangle  = \vert 0 \rangle, \quad i = k_{\rm max} + 1, k_{\rm max} + 2, \cdots, N,
$$
so that the Bargmann function takes the form
$$
\psi(z) = d_{k_{\rm max}} \prod_{i=1}^{k_{\rm max}} \sqrt{1 + {\bar \omega_i} \omega_i} 
                          \prod_{j=1}^{N} \langle {\bar z} \vert \omega_j \rangle. 
$$
Since the representation $\psi_d \mapsto \psi$ is unique up to permutations of the $\vert \omega_i \rangle$, the Bargmann function can be rewritten as 
$$
\psi(z) = 
d_{k_{\rm max}} \prod_{i=1}^{k_{\rm max}} \sqrt{1 + {\bar \omega_i} \omega_i}\frac{1}{N!} \sum_{\sigma \in S_N} 
                \prod_{j=1}^{N} \langle {\bar z} \vert \omega_{\sigma(j)} \rangle
$$
or alternatively as
\begin{eqnarray}
\psi(z) = {\cal N}_d \sum_{\sigma \in S_N} \langle N : {\bar z} \vert \omega_{\sigma(1)} \omega_{\sigma(2)} \cdots \omega_{\sigma(N)} \rangle, 
\label{expression-psi2}
\end{eqnarray}
where the normalization constant ${\cal N}_d$ is given by
$$ 
{\cal N}_d = \frac{1}{N!} d_{k_{\rm max}} \prod_{i=1}^{k_{\rm max}} \sqrt{1 + {\bar \omega_i} \omega_i}.
$$
Comparing Eqs.~(\ref{expression-psi1}) and (\ref{expression-psi2}), we find that the state $\vert \psi_d \rangle$ can be expressed as
\begin{eqnarray}
\vert \psi_d \rangle = {\cal N}_d \sum_{\sigma \in S_N} 
                      \sigma (\vert \omega_1 \rangle \otimes 
											        \vert \omega_2 \rangle \otimes 
															\cdots \otimes 
															\vert \omega_N \rangle)
\label{psidenomega}
\end{eqnarray}
in terms of the $k_{\rm max}$ zeros $\omega_i$ for $i = 1, 2, \cdots, k_{\rm max}$ of the Bargmann function $\psi$ and of their extension 
$\omega_{k_{\rm max + 1}} = 
 \omega_{k_{\rm max + 2}} = 
                   \cdots = 
\omega_{N} =0$. 

\subsection*{Expression of $\vert \psi_d \rangle$ in terms of the Majorana stars}

We note that the states $\vert \omega_i \rangle$ with $i = 1, 2, \cdots, k_{\rm max}$ can be written in terms of the coherent states 
$\vert z_i \rangle \equiv \vert 1 : z_i \rangle$ by putting  
\begin{eqnarray*}
z_i = \cases{-\frac{1}{\omega_i} \ {\rm if} \ i = 1, 2, \cdots, k_{\rm max}, \cr \cr
             0                   \ {\rm if} \ i = k_{\rm max}+1, k_{\rm max}+2, \cdots, N. 
}
\end{eqnarray*}
We verify that
$$ 
\vert \omega_i \rangle = \vert z_i \rangle, \quad i = 1, 2, \cdots, N
$$
up to irrelevant phase factors. Hence, the symmetric qudit state $\vert \psi_d \rangle$ given by (\ref{psidenomega}) can be expressed as 
\begin{eqnarray}
\vert \psi_d \rangle = {\cal N}_d  \sum_{\sigma \in S_N} \sigma 
(\vert z_1 \rangle \otimes 
 \vert z_2 \rangle \otimes
 \cdots \otimes
 \vert z_N \rangle) 
\label{psidenomegabis}
\end{eqnarray}
in terms of the coherent states $\vert z_i \rangle$. Equation (\ref{psidenomegabis}) is identical to (\ref{formcoherentstateN}): we thus recover 
Eq.~(\ref{formcoherentstateN}).

\subsection*{Equation satisfied by the Majorana stars}

The zeros $\omega_i$ of the Bargmann function (\ref{psi(z)enP(z)}) satisfy ${\cal P}(\omega_i) = 0$. From Eq.~(\ref{detailed expression of P(z)}), we thus get
\begin{eqnarray*}
\sum_{k=0}^{k_{\rm max}} \sqrt{\frac{N!}{k!(N-k)!}} c_k \omega_i^{k}  = 0 \ \Rightarrow \
\sum_{k=0}^{N}           \sqrt{\frac{N!}{k!(N-k)!}} c_k \omega_i^{k}  = 0
\end{eqnarray*}
or in terms of the $z_i$
$$
\sum_{k=0}^{k_{\rm max}} (-1)^k \sqrt{\frac{N!}{k!(N-k)!}} c_k z_{i}^{N-k}  = 0 \ \Rightarrow \ 
\sum_{k=0}^{N}           (-1)^k \sqrt{\frac{N!}{k!(N-k)!}} c_k z_{i}^{N-k}  = 0
$$
in agreement with Eq.~(\ref{d'Alembert}). 

\section{Concluding remarks}

In this paper we discussed the role of a specific generalized Weyl-Heisenberg algebra in the algebraic structure of qubits and qudits. The use of this generalized Weyl-Heisenberg algebra is based on the fact that qubits are neither fermions nor bosons. Indeed, in the standard theoretical approach of quantum information, a qubit is a vector in a two-dimensional Hilbert space as for fermions and the Hilbert space of a multiqubit system has a tensor product structure like for bosons. In this respect, the commutation rules of the raising and lowering operators for qubits are not specified by relations of bosonic type or of fermionic type. 
\\

By using a collection of $N=d-1$ qubits, we gave a realization of the $d$-dimensional representation space of the generalized Weyl-Heisenbeg algebra. In particular, we demonstrated that the vectors of this representation space coincide with the Dicke states. These states are of special interest for describing multiqubit quantum systems possessing exchange symmetry. Another advantage of this algebraic description via the generalized Weyl-Heisenberg algebra concerns the separability of multiqubit states invariant under permutations. Hence, starting from the decomposition of Dicke states, we investigated the condition for the separability of symmetric qudits made of $N = d-1$ qubit states. Our results show that exchange symmetry implies that the superposition of Dicke states are globally entangled unless they are fully separable and coincide with the coherent states, in the Perelomov sense, associated with the generalized Weyl-Heisenberg algebra. 
\\

In the Majorana description of a symmetric qudit state in terms of symmetrized tensor products of $N = d-1$ qubits, we introduced a parameter $P_d$ connected to the permanent of the matrix characterizing the overlap between the $N$ qubits. This parameter provides us with a quantitative measure of the entanglement for the qudit arising from $N$ qubits. This was illustrated in the special case $d = 3$, for which the parameter $P_d$ constitutes an alternative to the Wootters concurrence $C$ for $N=2$ qubits. Therefore, we propose that $P_d$ be called {\em perma-concurrence} as a contraction of {\rm permanent} and {\rm concurrence}. Other examples of $P_d$ were given for $d = 4$ and $5$. The results highlight the interest of the perma-concurrence $P_d$ for measuring the entanglement of a symmetric qudit state developed in terms of tensor products of qubit coherent states.
\\

In the annexe, we further investigated the formalism of qubit coherent states to describe qudit states in the Fock-Hilbert space corresponding to the generalized 
Weyl-Heisenberg algebra. More precisely, we used the Fock-Bargmann representation for describing any symmetric qudit constructed from $N = d - 1$ qubits with the help of an analytic function, the so-called Bargmann function. The zeros of the Bargmann function were related to the Majorana stars which provide an alternative way to describe Fock-Hilbert states as tensor products of qubit coherent states labeled by complex variables, namely, Majorana stars on the Bloch sphere. 
\\

Recently, new entropic and information inequalities for one qudit, which differs from a multiqubit system, have been developed \cite{Manko}. Therefore, it will be a challenge to ask whether the qudit picture proposed in this paper can be adapted in terms of linear combinations of Dicke states.  
\\

To close this paper, note that it might be interesting to introduce Dicke states in the construction of the so-called mutually unbiased bases used in quantum information. This approach, feasible in view of the connection between mutually unbiased bases and angular momentum states \cite{KiblerLicata}, could be the object of a future work.

\vspace{2cm}

{\bf Author Contributions}

The authors equally contributed to the paper. 
\\

{\bf Conflicts of Interest}

The authors declare no conflict of interest.

\end{document}